# Contact resistance and interfacial engineering: Advances in high-performance 2D-TMD based devices

Xiongfang Liu [1], Kaijian Xing [2,3], Chi Sin Tang [*,4], Shuo Sun [1], Pan Chen [1], Dong-Chen Qi [5,6], Mark B. H. Breese [4,7], Michael S. Fuhrer [2,5], Andrew T. S. Wee [*,8], Xinmao Yin[*,1]

[1] Shanghai Key Laboratory of High Temperature Superconductors, Institute for Quantum Science and Technology, Department of Physics, Shanghai University, Shanghai 200444, China

[2] School of Physics and Astronomy, Monash University, Clayton, Victoria 3800, Australia

[3] School of Science, RMIT University, Melbourne, VIC 3001, Australia

[4] Singapore Synchrotron Light Source (SSLS), National University of Singapore, Singapore 117603, Singapore

[5] Australian Research Council Centre of Excellence in Future Low-Energy Electronics Technologies (FLEET), Monash University, Clayton, Victoria 3800, Australia

[6] Centre for Materials Science, Queensland University of Technology, Brisbane, Queensland 4001, Australia

[7] School of Chemistry and Physics, Queensland University of Technology, Brisbane, Queensland 4001, Australia

[8] Department of Physics, Faculty of Science, National University of Singapore, Singapore 117542, Singapore

**\*** To whom correspondence should be addressed:
slscst@nus.edu.sg (C.S.T.);
phyweets@nus.edu.sg (A. T. S. W.)
yinxinmao@shu.edu.cn (X.Y.)

**Abstract:**

The development of advanced electronic devices is contingent upon sustainable material development and pioneering research breakthroughs. Traditional semiconductor-based electronic technology faces constraints in material thickness scaling and energy efficiency. Atomically thin two-dimensional (2D) transition metal dichalcogenides (TMDs) have emerged as promising candidates for next-generation nanoelectronics and optoelectronic applications, boasting high electron mobility, mechanical strength, and a customizable band gap. Despite these merits, the Fermi level pinning effect introduces uncontrollable Schottky barriers at metal–2D-TMD contacts, challenging prediction through the Schottky-Mott rule. These barriers fundamentally lead to elevated contact resistance and limited current-delivery capability, impeding the enhancement of 2D-TMD transistor and integrated circuit properties. In this review, we succinctly outline the Fermi pinning effect mechanism and peculiar contact resistance behavior at metal/2D-TMD interfaces. Subsequently, highlights on the recent advances in overcoming contact resistance in 2D-TMDs devices, encompassing interface interaction and hybridization, van der Waals (vdW) contacts, prefabricated metal transfer and charge-transfer doping will be addressed. Finally, the discussion extends to challenges and offers insights into future developmental prospects.

**1. Introduction:**

Semiconductor devices are the backbone of modern electronics, driving innovation and technological advancement across various sectors. Conventional semiconductors, such as elemental semiconductor like Si, Ge and compound semiconductors like GaAs, InP, have played a pivotal role in shaping the landscape of electronic devices [1-3]. However, as the demand for higher performance and novel functionalities continues to soar, the limitations of traditional semiconductors have become increasingly apparent. These limitations, including constraints on device miniaturization, power consumption, and operating speed, underscore the need for exploring alternative materials and device architectures.

Amidst this quest for innovation, two-dimensional (2D) transition metal dichalcogenides (TMDs) have emerged as a beacon of hope. These atomically thin materials possess remarkable properties, including high carrier mobility, flexibility, and optical tunability, making them highly attractive for next-generation advanced devices, such as field effect transistors, solar cells, biosensors, absorbers, and photodetectors [4-7]. Unlike their bulk counterparts, 2D-TMDs exhibit unique quantum confinement effects and enhanced surface-to-volume ratios, offering unprecedented opportunities for device miniaturization and performance enhancement. Critical to the realization of high-performance advanced devices utilizing 2D-TMDs is the establishment of efficient metal contacts. These contacts serve as the conduits for current input and output processes, fundamentally influencing the device's overall performance [8-10]. However, a formidable challenge in this realm is the presence of the Schottky barrier, a barrier to charge carrier transport arising from Fermi level pinning effects at the metal-2D-TMD interface [11,12].

During the fabrication of 2D-TMD-based devices, processes such as metal deposition can introduce severe lattice defects and interfacial damage, leading to unavoidable Fermi level pinning effect and the formation of high-resistance contacts. Overcoming this challenge is paramount for unlocking the full potential of 2D-TMD devices, as high contact resistance can severely limit device performance metrics such as speed, power consumption, and reliability [13,14]. In this context, addressing the contact resistance problem in 2D-TMD materials emerges as a critical area of focus for device engineers and material scientists alike. The ability to establish low-resistance metal contacts not only enhances the efficiency of charge carrier transport but also facilitates the realization of advanced device functionalities.

A number of reviews offer a good summary in contact resistance and interfacial engineering of TMDs [11,15-19]. Both Schulman et al. [17] and Allain et al. [15] have provided very comprehensive treatments concerning the fundamental understand for metal-TMDs contact and interfacial dynamics while presenting interfacial engineering strategies in overcoming contact resistances. Over the course of 5 years, significant progress have been made to minimize contact resistances via the development of new interfacial engineering techniques and novel insights including interface hybridization (metallization of TMDs [20], creating edge contacts [9,21], and inserting buffer layers [22]), vdW contacts [10], prefabricated metal transfer [23], and charge-transfer doping [24]. While reviews published within the recent two years particularly by Wang et al. [11] and Ma et al. [19] have been insightful in keeping readers abreast with rapid progresses related to contact strategies, mechanisms, and a wide range of developed 2D electronics. it is necessary to consider the nuanced understanding and strategic implementation of these distinct approaches towards the realization of high-performance 2D-TMD-based electronic devices to be utilized in emerging disciplines including spintronic applications, catalytic, energy storage, neuromorphic computing systems (Figure 1) [16,25-27]. At this juncture, new emerging disciplines related to spintronic devices and optoelectronic devices especially *p*-type electronic devices are prompted by the burgeoning drive for metal/2D-TMDs interfaces engineering has prompted significant interests in the utilization of such low-dimensional systems [28-30]. While extensive work has been performed on metal/2D-TMDs interfaces especially with their vast potential for high-performance 2D-TMD based devices in the past 3 years, it is imperative to take into consideration the eventual goal of developing and integrating these fundamental techniques with advanced manufacturing processes amidst the rapid scientific advances of 2D-TMDs concerning the contact resistance and interfacial engineering.

In this review, we embark on a journey to unravel the mysteries surrounding the Fermi level pinning effect and its implications for contact resistance in TMD materials. By delving into the physical underpinnings of this phenomenon, we aim to provide a comprehensive understanding of the challenges posed by contact resistance in 2D-TMD-based devices. Furthermore, we meticulously examine a diverse array of interfacial engineering strategies aimed at circumventing these challenges, ranging from interface interaction and hybridization (including metallization of TMD, edge contact, insertion of buffer layers) to vdW contacts, prefabricated metal transfer and charge-transfer doping. Through our systematic analysis, we endeavor to shed light on the promising avenues for overcoming contact resistance barriers in 2D-TMD materials, paving the way for the development of high-performance electronic devices with unparalleled efficiency and functionality.

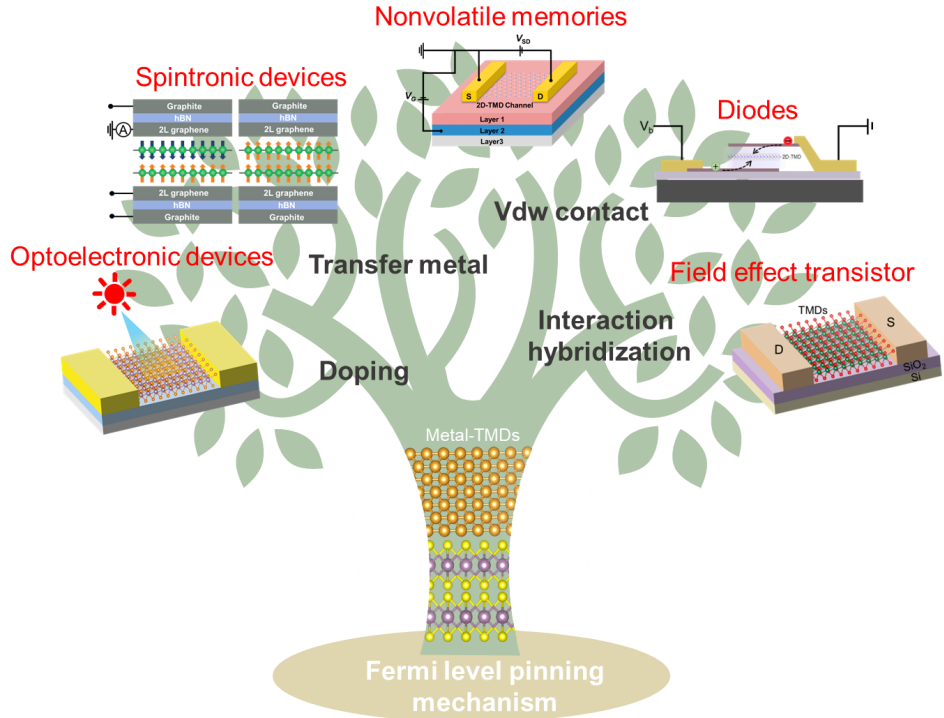

Figure 1. The accessibility of various efficient low contact resistance and interfacial engineering techniques has facilitated the utilization of low-dimensional transition metal dichalcogenide (TMD) materials in numerous electronic, optoelectronic, and spintronic applications.

## 2. Metal contacts with 2D-TMD materials

### 2.1 Metal contacts with ideal conventional semiconductor

Ohmic contact and Schottky contact are two main forms of contacts between metal and conventional semiconductor. Under ideal conditions, the corresponding current-voltage curve between the metal and the conventional semiconductor is a straight line, indicating that ohmic contact has negligibly low contact resistance between them. Nevertheless, in conventional metal/semiconductor interfaces, the equilibrium condition is no longer similar to that of bulk systems where an energy gap between valence and conduction bands is present. Instead, a continuum of localized states which are either occupied or empty are present. To achieve charge equilibrium at the interface and within the bandgap, localized states up to what is known as the *charge neutrality level*, $E_{CNL}$, are occupied and those above are empty. Consequently, the Fermi level must be at or near this value at the interface, $E_F = E_{CNL}$, so as to meet the equilibrium conditions [31]. The valence and conduction bands of the conventional semiconductor at the interface would therefore be fixed with the corresponding bending of the semiconducting bands at the interface (Figures 2a-b). This results in the formation of an interfacial depletion region where charges in the region are balanced by a minute adjustment of the Fermi level away from $E_{CNL}$ to ensure charge equilibrium at the semiconductor-metal interface. The height of the Schottky barrier is primarily affected by the work function of metal in ideal metal–conventional semiconductor contact [32,33]. Therefore, the value of Schottky barrier height (SBH) is the difference between work function of metal $q \cdot \phi_m$ and electron affinity of semiconductor $q \cdot \chi$. The built-in potential, $V_{bi}$, is the difference between work function of metal $q \cdot \phi_m$ and semiconductor $q \cdot \phi_s$. Hence, SBH and $V_{bi}$ can mathematically expressed as the following:

$$SBH = q \cdot \phi_m - q \cdot \chi \tag{1}$$

$$V_{bi} = q \cdot \phi_m - q \cdot \phi_s \tag{2}$$

It is instructive to note that a high and wide Schottky barrier is detrimental to device performance because charge depletion leads to anomalous high interfacial resistance. Under extreme scenarios, the entire interfacial layer may be depleted of charge and no conductivity can be registered [34,35].

The choice of metals with suitable work functions can minimize SBH at ideal metal/conventional semiconductor interfaces. Another important method of tuning interfacial charge density is by applying a bias voltage at both ends of the metal–conventional semiconductor heterojunction [36,37]. By Considering the contact between a metal and a n-type semiconductor where $q \cdot \phi_m > q \cdot \phi_s$, the addition of a forward bias can reduce the difference of the $V_{bi}$ and the width of depletion region, thereby increasing the injection of interfacial charges. Conversely, the reverse voltage will increase the difference between $V_{bi}$ and the depletion region width. This further increases the contact resistance (Figures 2c-d). However, the SBH of realistic metal/conventional semiconductor interfaces is not tunable by applying a bias voltage.

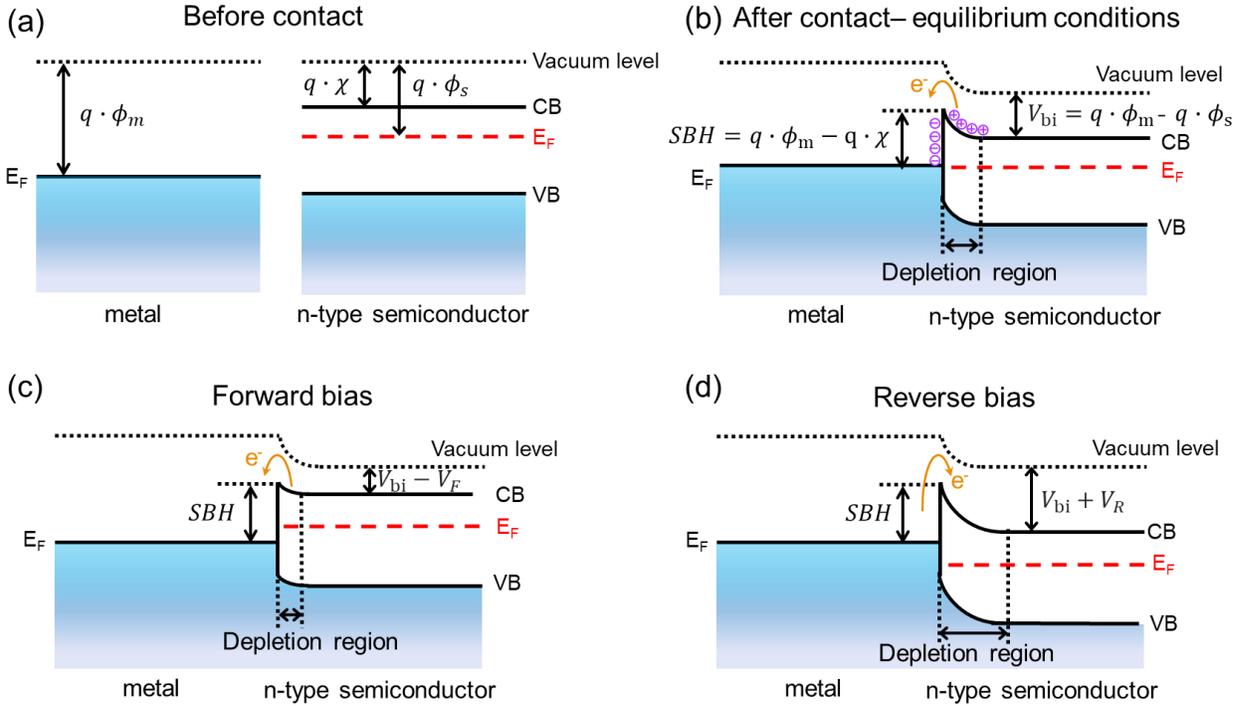

Figure 2. (a) Energy band diagram of independent metal and semiconductor (n-type) before contact. Of these, $q \cdot \phi_m > q \cdot \phi_s$. Energy band diagram of a Schottky barrier under (b) equilibrium conditions, (c) forward bias $V_F$ and (d) reverse bias $V_R$. Under forward bias, allowing large amounts of electrons to flow from the semiconductor to the metal.

## 2.2 Fermi level pinning

Compared to conventional semiconductors, selected 2D-TMDs offer several advantages including high carrier mobility, tunable band gap, large specific surface area, and mechanical flexibility [18,38]. These properties make them highly promising for a wide range of electronic devices, particularly field-effect transistors (FETs). Pivotal to the integration of 2D-TMDs into electronic-based applications is the crucial step of establishing reliable contacts between the 2D-TMDs and metal electrodes, connectors, and supporting substrates [8-10]. The structure and electrical properties of metal/2D-TMD interfaces profoundly influence the performance of nano-electronic devices. Achieving ohmic contact is essential for ensuring efficient metal–2D-TMD interactions in electronic devices. Given the exceptional properties of metal–2D-TMD contacts, there is a pressing need for a comprehensive understanding of their fundamental principles.

Unlike ideal metal–conventional semiconductor contacts, Fermi level pinning has a considerable influence on metal–2D-TMD contacts. As a result, the Schottky barrier between metals and TMD materials exhibits minimal sensitivity to changes in the metal work function, presenting challenges in achieving ohmic contact through adjustments to the metal work function to align with that of 2D-TMD materials [39]. While the negative effects of Fermi-level pinning are widely recognized, the origin of this phenomenon remains inconclusive. In a study involving mechanically exfoliated $MoS_2$, McDonnell *et al.* attributed Fermi-level pinning to the effects of surface defects and chemistry-related variations in $MoS_2$ stoichiometry [40]. This is confirmed in a separate comprehensive experimental study by Addou *et al.* which points to structural and metallic-like defects, chalcogen vacancies, and other elemental impurities as the chief causes of Fermi level pinning [41]. In separate experimental and computational studies, the presence of metallic nanostructures such as Au, Pd and Ag is found to have a direct influence on the interfacial and electronic properties of the metal-TMD contact, due to inhomogeneity and interfacial lattice strain [42-44].

Reports have also attributed the onset of Fermi-level pinning to the formation of surface states created by adsorbed contaminants, as further suggested by McDonnell *et al* [45]. Specifically, by comparing the formation of the Ti-$MoS_2$ interface under high vacuum (~$1\times10^{-6}$ mbar) and ultrahigh vacuum (~$1\times10^{-9}$ mbar) conditions, interfacial $TiO_2$ is formed. Metallic Ti can readily react with $MoS_2$ at the interface that leads to the formation of elemental Mo, $Mo_xS_y$, and $Ti_xS_y$ which in turn dominate the contact behavior with relatively high contact resistance [46]. Conversely, the deposition of $TiO_2$ does not react with the $MoS_2$ layer. Nevertheless, the unintentional introduction of $TiO_2$ contaminants leads to an increased thermal boundary resistance [47,48], which in turn will affect the interface contact properties.

In addition, it is noted that the material integration and device fabrication processes inevitably lead to interface chemical reaction, interdiffusion and defect-induced gap states [49] that can serve as a pool for electrons or holes that pins the Fermi level [49,50]. Such lattice defects and interfacial damage can take place due to radiation or ionization techniques where metal deposition processes take place with the bombardment of high-energy atoms and strong local heating to the contact region [13,14]. Such effects are commonly observed in III–V based semiconductors [51].

Using density functional theory (DFT) calculations, Gong et al. reported that the mechanisms of Fermi-level pinning at metal-MoS$_2$ contacts as unique to metal/2D-TMD interfaces are distinct from the pinning effect found at conventional metal-semiconductor junctions [52]. Fermi-level pinning at the metal–2D-TMD interfaces is a result of two main mechanisms. Namely, there is an onset of metal work function modification due to the formation of interfacial dipoles induced by charge redistribution. Computational results reveal that for metal substrates such as Al with deeper d-band than the *s*- and *p*-bands, there is an increase in effective metal work function, while in other substrates such as Ag, Au, Ir, Pd and Pt, the work function decreases. The results are generally consistent with a separate computational study involving the adsorption of metals on MoS$_2$ with a wide range of Group I—IV metals, with a work function reduction observed in the MoS$_2$-metal system [53]. Nevertheless, a separate study shows that the charge-transfer direction which in turn affects the dipole direction is dependent on the interfacial configuration, as shown in the case of Pt nanoparticles adsorbed on MoS$_2$ [54].

The second mechanism proposed by Gong et al. is the formation of gap states, distinct from the metal-induced gap states (MIGS) that comprises mainly Mo *d*-orbital features due to the weakened intralayer S–Mo bonding with interfacial metal–S interaction. However, the scanning tunneling microscopic (STM) study shown in Figure 3a-b by Kerelsky et al. has shown otherwise, where the formation of the electronic properties within 2 nm of the MoS$_2$-metal interface is dominated by the local density of states, consistent with that of MIGS (Figure 3c) [55]. The argument against the second mechanism suggested by Gong et al. is proposed in a separate DFT study by Chen et al. involving monolayer-MoS$_2$ adsorbed on Ir, Pd and Ru substrates where partial Fermi level pinning and charge transfer from the metal substrate to monolayer-MoS$_2$ is registered (Figure 3d-e) [56]. While there are suggestions of the MIGS mechanism present in this study, the monolayer-MoS$_2$ thickness restricts the extent of Fermi-level pinning due to spatial limits to the density of interfacial states formed [57].

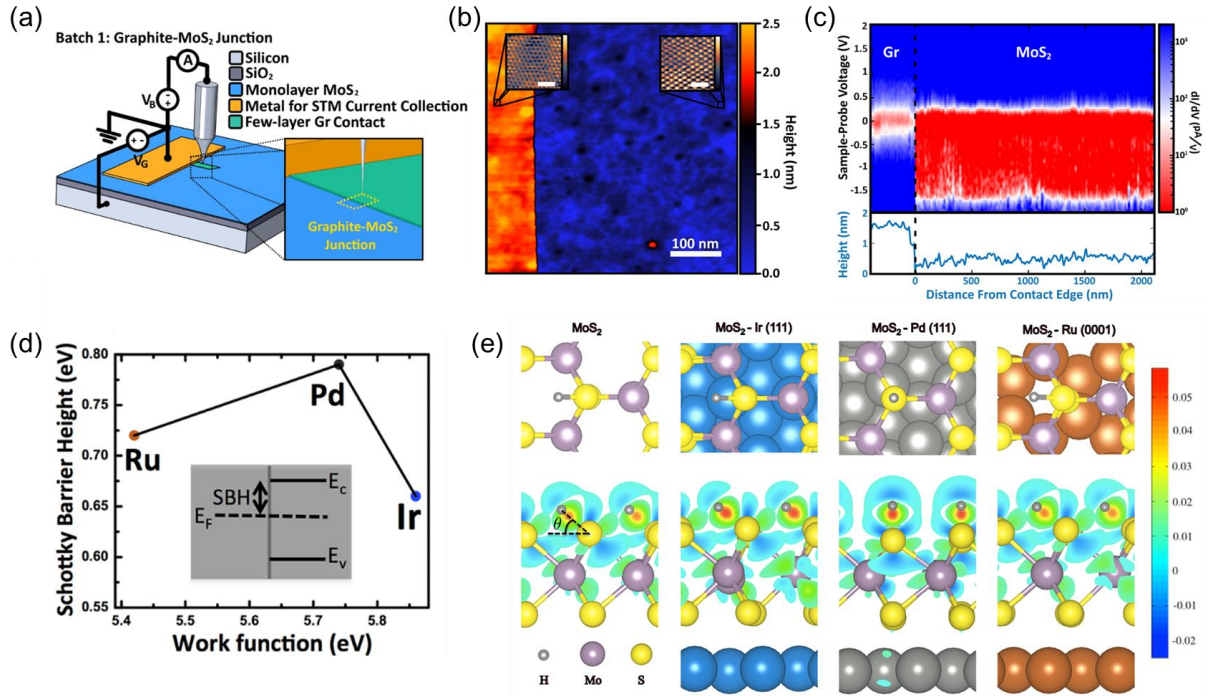

Figure 3. Experimental setup for the graphite-MoS$_2$ junction. (a) Device schematic, (b) STM topographic image at the edge of the graphite electrode above the monolayer-MoS$_2$, with its (c) color map of a short (~9 nm) dI/dV line profile taken across the sample alongside the approximate location of the contact edge at the bottom of the figure. (d) Schottky barrier heights of the metal-MoS$_2$ interface that is dependent on the respective work function, and (e) top and cross-section view indicating charge-transfer density of free-standing MoS$_2$, MoS$_2$/Ir, MoS$_2$/Pd and MoS$_2$/Ru. Panel (a)–(c) adapted from ref. [55] with permission from American Chemical Society, copyright 2017. Panels (d) reproduced from ref. [56] with permission from American Chemical Society, copyright 2013.

## 2.3 Fundamentals of metal-TMD contacts

The structural and electronic behaviour at the 2D-TMDs and metal interfaces have significant influences on the performance of 2D electronic devices [58]. The types of metal–2D-TMD contacts include covalent and the vdW interfaces, as suggested by previous theoretical investigations in systems such as MoS$_2$/Ti and MoS$_2$/Au [59,60]. However, the onset of Fermi-level pinning at the interfaces undermines the performance of heterostructures formed by the deposition or lithography of metals onto 2D-TMD surfaces [15,39,52]. Atomic-scale characterization studies conducted by scanning transmission electron microscopy (STEM) at the metal–2D-TMD interfaces provide important evidence that the degradation in device performance is associated with the damage of the 2D-TMD in the contact layer due to the invasion of interfacial metal atoms [39,61,62]. To circumvent this problem, interface engineering processes in the form of inserting additional vdW layers such as graphene [63] or hexagonal boron nitride (hBN) [64] have been explored. The mechanical transfer of metallic films [61] and the infusion of sub-monolayer indium [10] have also been investigated in the quest for ideal vdW contacts at metal-TMD interfaces.

It has been suggested in some studies that metal-TMD contacts such as the $MoS_2$/Au system are of vdW instead of covalent nature. With the removal of the Au substrate, the optical and electronic properties of monolayer $MoS_2$ resemble those of the semiconducting 1H-phase that have been exfoliated directly onto insulating substrates [65-67]. While this implies the reversibility in the interfacial effects of TMD layers endowed by the metallic substrates [66,68], concerted efforts have to be put into the understanding of the interaction at the metal-TMD contact and the effects of these interactions.

To provide further understanding of the nature of TMD-metal interactions, an *ab initio* DFT study was conducted by Popov *et al*. to investigate the effects of geometry, bonding, and electronic structures of the contact region [59]. Results show that unlike chemically non-saturated sulfur that forms favorable thiol bonds with Au, the S-atoms in $MoS_2$ are fully saturated and do not bond strongly with the Au atoms. This results in a large separation between the S-atoms in the $MoS_2$ layer and the closest Au neighboring atom–longer than the Au-S covalent length. This distance effectively suppresses any efficient wave function overlap, thus forming a physical separations and tunnel barrier between the Au-$MoS_2$ interface, making it inefficient for electron injection. In the case of the Ti-$MoS_2$ contact, the distance between the S-atoms and closest Ti atoms is significantly smaller than at the Au-$MoS_2$ interface. In principle, this makes the $MoS_2$/Ti a more favourable system due to the higher density of states at the Fermi level with a higher interfacial carrier density. However, experimental studies still indicate limited charge injection at the Ti-TMD contacts [69]. Despite being an inefficient contact for electron injection [70], FET systems with Au contacts generally have the lowest contact resistance due to the weak Fermi level pinning effect at the Au-TMD contacts [71].

## 3. Interfacial Effects & Contact Engineering: Recent Progress in Overcoming Contact Resistance

The Fermi level pinning effect, inherent in metal–2D-TMD contacts, poses challenges for achieving precise band alignment between 2D-TMDs and metals, resulting in the formation of Schottky barriers along the energy band's vertical direction. These high Schottky barriers present significant hurdles in achieving optimal ohmic contacts. Consequently, ongoing efforts focus on exploring various engineering strategies to overcome this challenge, aiming to mitigate interfacial Fermi level pinning and enhance device performance in metal–2D-TMD devices. While the experimentally induced factors can lead to the onset of contact resistance and Fermi-level pinning, these adverse effects could possibly be alleviated through metal-deposition and lithographic processes [12,72]. However, the underlying interfacial interaction processes, such as vdW forces and hybridization between the metal and TMD system, stem from the intrinsic properties of the materials. Thus, addressing these processes presents a more fundamental challenge to be resolved.

### 3.1 Interfacial Interaction and Hybridization

Interfacial hybridization engineering (as illustrated in Figure 4) significantly influences Fermi level pinning. Leveraging interface interactions and atomic-level hybridization between 2D-TMDs and metals represents an effective strategy to harness the Fermi level pinning effect for the improvement of ohmic contacts [12]. Interfacial hybridization engineering encompasses two primary strategies. Firstly, enhancing the interaction between 2D-TMDs and metals directly modulates interfacial hybridization. Strengthening this interaction can effectively mitigate the detrimental Fermi level pinning effect by promoting the metallization of 2D-TMDs. Secondly, weakening the interaction between metals and 2D-TMDs to achieve a weaker interfacial hybridization reduces the Fermi level pinning effect by minimizing the presence of gap states.

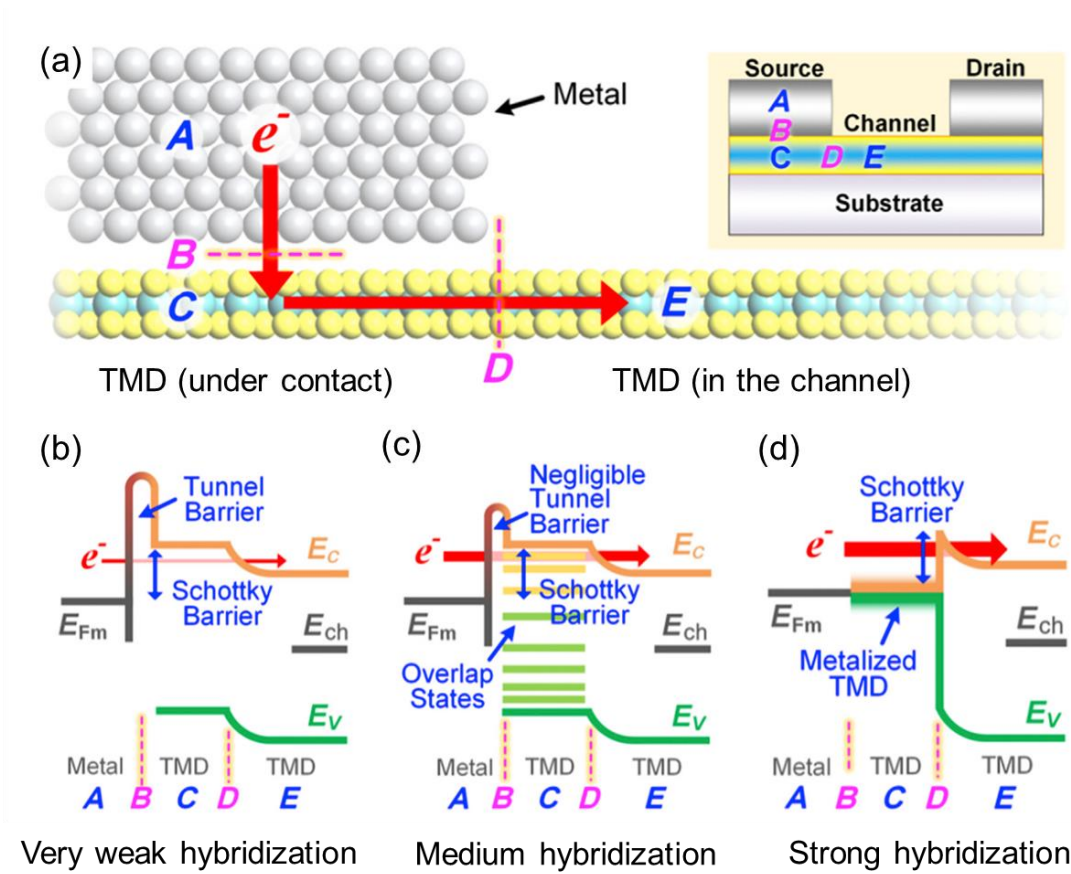

Figure 4. (a) An illustration of a typical metal-MoS$_2$ contact's cross-section (n-type top contact). The three zones are indicated by the letters A, C, and E, and the two interfaces that divide them are B and D. The electron injection pathway from contact metal (A) to the MoS$_2$ channel (E) is shown by red arrows. A typical backgated FET's channel area and source and drain connections are seen in the inset. (b)–(d) The energy band diagrams of three types of interfacial hybridization. (b) Band diagrams showing a vdW tunnel barrier caused by extremely weak hybridization. (c) Band diagrams for medium hybridization induced interfacial states. (d) Band diagrams presenting the metallization of TMDs due to strong hybridization. Panels (a)-(d) reproduced from ref. [60] with permission from American Chemical Society, copyright 2014.

### 3.1.1 Strong interfacial hybridization

The strength of interfacial hybridization is primarily assessed by key criteria including orbital overlaps, Schottky barrier, and tunnel barrier within the bandgap of 2D-TMDs [60]. Typically, smaller atomic sizes and the utilization of the same metal element in both 2D-TMDs and metals result in reduced physical separations, specifically the nearest core-to-core distance between metal atoms and chalcogenide atoms in the z-direction. These diminished physical separations may yield exceptionally thin tunnel barriers and robust orbital overlap [60]. The greater population and large size of the atomic orbitals near $E_F$ have considerable influence on real-space overlap of metal and 2D-TMDs orbital which leads to strong band hybridization in the metal–2D-TMD contacts [20]. This strong orbital overlap results in a strong distortion of 2D-TMDs band structures which result in overlapping states in the original band gaps. Thus, the electronic properties of the 2D-TMDs under contact are significantly modified, resulting in strong metallization to 2D-TMDs and a strongly depressed Schottky barrier [73]. Disappearing Schottky barrier and eliminating Fermi level pinning effect at the metal/2D-TMD interfaces are the ideal ohmic contact targets.

Experimentally, Li *et al.* reported a significant reduction in contact resistance to the quantum limit, achieved through band hybridization between monolayer-$MoS_2$ and semi-metallic Sb ($01\bar{1}2$) [20]. The strong band hybridization in the $MoS_2$–Sb ($01\bar{1}2$) contact leads to charge transfer at the interfaces (Figure 5a) and effectively pulls the $MoS_2$ conduction band down through the $E_F$. Using a transfer-length-method (TLM) device, the excellent contact resistance, $R_c$, of the field-effect transistors consisting of $MoS_2$–Sb ($01\bar{1}2$) is derived to be 42 Ω·μm. The on-state current is registered at 1.23 mA·μm$^{-1}$ with an on/off ratio over $10^8$ and intrinsic gate delay of 74 fs. All values are superior to previous reports. These experimental findings have been substantiated by separate DFT results reported by Kim, *et al* which indicated the formation of hybridization between $MoS_2$ and In conduction band edge states as the microscopic origins of ohmic charge injection between $MoS_2$ and metal contacts (Figure 5b) [74].

Doping with chalcogen vacancies and interstitials represents another promising set of techniques for engineering interactions between metals and 2D-TMDs, with chalcogen vacancies boosting the orbital overlaps of 2D-TMDs with all metals. This is because such vacancies reduce the interfacial physical distance at the metal−2D-TMDs contact, which in turn can suppress the tunnel barrier and contact resistance. Interestingly, the presence of chalcogen vacancies leads to the conversion of TMDs−Pd contacts from an n-type to a p-type interface due to the onset of interfacial charge transfer [71]. In the case of interstitial doping process, the doped chalcogen atom interacts with metallic atoms, thereby reducing the bonding distance and improving charge injection at certain

metal−2D-TMD interfaces [71]. This method effectively reduces contact resistance for $MoS_2$ with Cr and Pd, $MoSe_2$ with Au, $WS_2$ with Au, and $WSe_2$ with Cr and Ni [75,76].

Strong interfacial hybridization can also be achieved by the edge contact method [21]. The physical separation of edge contact between the TMDs and the metal is much smaller than the top contact. Smaller physical separations and the presence of dangling bond at 2D-TMDs edge cause stronger hybridization at the edge contact interface [60,77]. This can enhance carrier injection and have a narrower tunnel barrier compared to top contact. Wang *et al.* reported the fabrication of one-dimensional edge contacts (Figures 5c-d) [78]. In this case, 3D metal electrodes are connected to a single-layer graphene via a metalized 1D graphene edge [79-81]. Even though the carrier injection is confined to the 1D atomic edge of the graphene layer, the contact resistance can go as low as 100 Ω·μm. However, the edge contacts method is difficult to achieve in technique due to its small contact area. Based on this, Song, *et al* have proposed a new technique—through using thermally stable 2D metal [9], $PtTe_2$ as an epitaxial template enables the lateral growth of monolayer $MoS_2$ to fabricate low-resistance edge contacts (Figure 5e).

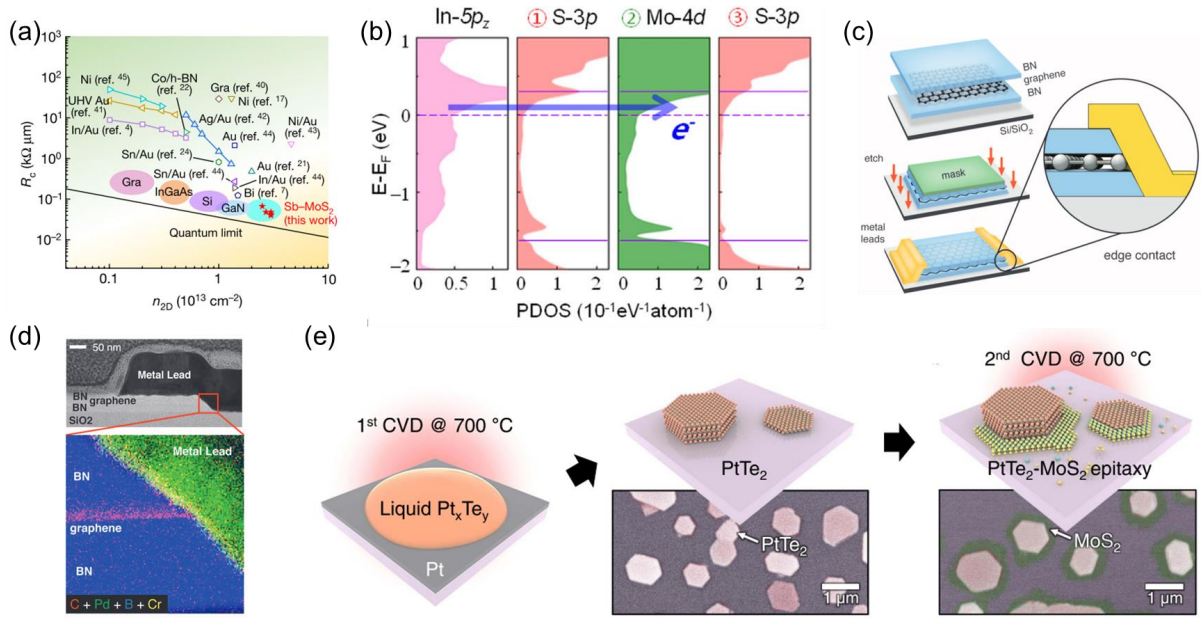

Figure 5. (a) Schematic representation depicting the lowest contact resistance ($R_c$) of 42 Ω μm at carrier concentration ($n_{2D}$) of $3\times10^{13}$ $cm^{-2}$, which is very approaching to quantum limit. (b) These four patterning are the projected density of states of In-$5p_z$, S-$3p$, Mo-$4d$ and S-$3p$ orbitals, respectively. The thick arrow represents electrons injecting from In into these $MoS_2$ in-gap states.

(c) Schematic of the edge-contact fabrication process, and (d) above: the high-resolution STEM image of the edge-contact geometry. Below: False-color EELS map of the graphene edge-metal interface. (e) $MoS_2$-$PtTe_2$ lateral heterostructure formation via two-step growth. The first step is growth of $PtTe_2$ flakes at a growth temperature of 700 °C by chemical vapor deposition (CVD) method. The second growth process is fabricating $MoS_2$ along the edge of $PtTe_2$ at a growth temperature of 700 °C by CVD method. The insets are scanning electron microscopy (SEM) pattens of product of each step. Panels (a) reproduced from ref. [20] with permission from Springer Nature Limited, copyright 2023. Panels (b) reproduced from ref. [74] with permission from Springer Nature Limited, copyright 2021. Panels (c)-(d) reproduced from ref. [78] with permission from the American Association for the Advancement of Science, copyright 2013. Panels (e) reproduced from ref. [9] with permission from Springer Nature Limited, copyright 2022.

### 3.1.2 Moderate interfacial hybridization

In contrast to the contact engineering based on strong metal–2D-TMD hybridization mentioned in the previous section 3.1.1, another method to eliminate the Fermi level pinning effect is to weaken the interfacial hybridization [16]. The moderate interactions between metals and 2D-TMDs can induce MIGS that pin the Fermi level in the band gap of 2D-TMDs and increase the SBH [82]. The extent of Fermi level pinning could essentially be alleviated by reducing the interfacial density of states by spatially separating the materials. As proposed by Gupta *et al.* in an experimental and theoretical study, spatial separation can be introduced via the addition of an intermediary contact that comprises an ultrathin insulating layer between the metal and 2D-TMDs materials [83]. This technique could attenuate the propagation of electron wave function from the metal layer, thereby reducing the formation of the interfacial gap states which lead to Fermi level pinning. The addition of an insulating layer leads to the formation of dipole moments at the insulator/2D-TMDs interfaces which reduces the SBH. However, achieving an optimal insulator thickness is crucial, as the introduction of this insulating layer creates a tunnel barrier that hinders carrier injection between the metal and TMD layers.

Recent advances have highlighted the effectiveness of ultrathin interlayers in modifying and improving the interfacial properties between 2D-TMDs and metal electrodes. For instance, DFT results by Lizzit *et al.* reported the increment in metal–2D-TMD distance leads to the effective suppression of the formation of MIGS and the SBH (Figure 6a) [82]. An alternative strategy proposed by Andrews *et al.* which implements the addition of an ultrathin 2D-TMDs ($MoSe_2$/$WSe_2$) as contact interlayers in $MoS_2$-Ti field-effect transistors can significantly reduce the SBH and contact resistivity [22]. This in turn leads to a significant enhancement in device performance (Figure 6b). Jang et al. demonstrated a metal−interlayer−2D-TMD contact configuration, incorporating an ultrathin ZnO interlayer between $MoS_2$ and the metal electrode (see Figure 6c) [84]. This approach effectively reduces interface disorder by depinning the Fermi level and enhances interaction at the interlayer/metal interfaces, thereby reducing the Schottky barrier and contact resistance. Meanwhile, the insertion of a SiC interlayer has also been proposed to counter the adverse effects of strong Fermi-level pinning [85]. The outcome of this strategy has proven to be promising with a substantial increase in the average charge value at the interface between SiC and $MoS_2$ layers, which in turn leads to a reduction in e the interfacial contact resistance.

While MIGS can arise due to moderate hybridization between 2D-TMDs and metals, this effect can be mitigated by the substitution of metals with semi-metallic alternatives. Based on this principle, Shen *et al.* have demonstrated successful contact between 2D-TMDs and semimetallic Bi, effectively suppressing MIGS and inducing the spontaneous formation of degenerate states in the TMDs upon contact with Bi (see Figure 6d) [57]. Through this approach, zero SBH and contact resistance of 123 Ω·μm and an on-state current density of 1,135 μA·μm$^{-1}$ on $MoS_2$–Bi field-effect transistors have been achieved.

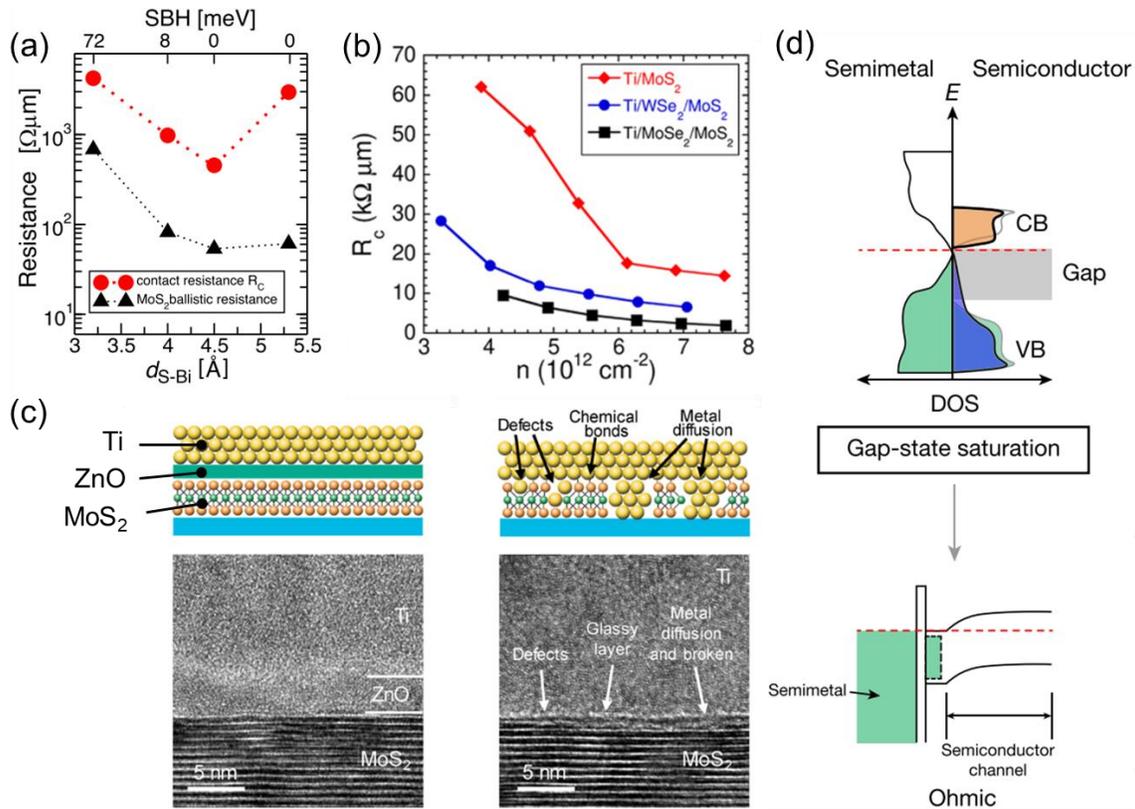

Figure 6. (a) Schottky barrier height (black points) and contact resistance (red points) for different distances $d_{S-Bi}$ in Bi−MoS$_2$ contact. (b) Comparison of extracted contact resistance for Ti/MoS$_2$, Ti/WSe$_2$/MoS$_2$, Ti/MoSe$_2$/MoS$_2$ contact at different carrier concentrates. (c) Cross-sectional schematic and TEM image shows that Ti/ZnO/MoS$_2$ contact (left) possesses cleaner metal−semiconductor interfaces than Ti/MoS2 contact (right). (d) low DOS of semimetal Bi at the Fermi level causes strong suppression of MIGSs and degenerate states spontaneously formed in the TMDs. Panels (a) reproduced from ref. [82] with permission from American Chemical Society, copyright 2022. Panels (b) reproduced from ref. [22] with permission from American Chemical Society, copyright 2020. Panels (c) reproduced from ref. [84] with permission from American Chemical Society, copyright 2020. Panels (d) reproduced from ref. [57] with permission from Springer Nature Limited, copyright 2021.

Based on the series of strategies and reported studies presented in this section, it has been clearly demonstrated that the combination of metals with strong atomic hybridization capabilities with 2D-TMDs have been demonstrated as good candidates for contact materials. Meanwhile, achieving reduced contact resistance through moderate atomic hybridization between metals and

2D-TMDs often necessitates additional procedures, such as the insertion of a buffer layer or the substitution of the metallic contacts with a semi-metallic material. Consequently, both strong and reduced hybridization pathways offer effective strategies for achieving optimal ohmic contacts.

**3.2 vdW Contact**

While it remains challenging to produce good vdW contacts between 2D-TMDs and 3D metal substrates, the formation of vdW contacts by multilayered TMDs with exfoliation of graphene [63,86] and metals [61] come with greater success due to the ability to produce defect-free and clean vdW interfaces. Specifically, Wang *et al*. reported an ultraclean vdW contact between monolayer-$MoS_2$ and a 10nm indium metal capped with a 100 nm gold electrode (In/Au) as displayed in Figure 7a where the ultraclean vdW contact has been confirmed by annular dark field scanning transmission electron microscopy and X-ray photoelectron spectroscopy [10]. Apart from displaying the lowest contact resistance compared to other electrode materials (Figures 7b-c), the linear output characteristic of the FET system also suggests the absence of a Schottky barrier (Figure 7d) [10]. Similar good contacts have also been achieved with other CVD-grown 2D-TMDs including $NbS_2$ (Figure 7e), $WS_2$ (Figure 7f) and $WSe_2$ (Figures 7g-h) with the deposition of the In/Au alloy contacts. These results suggest the formation of defect-free and ultraclean vdW interfaces via the use of the In/Au contacts.

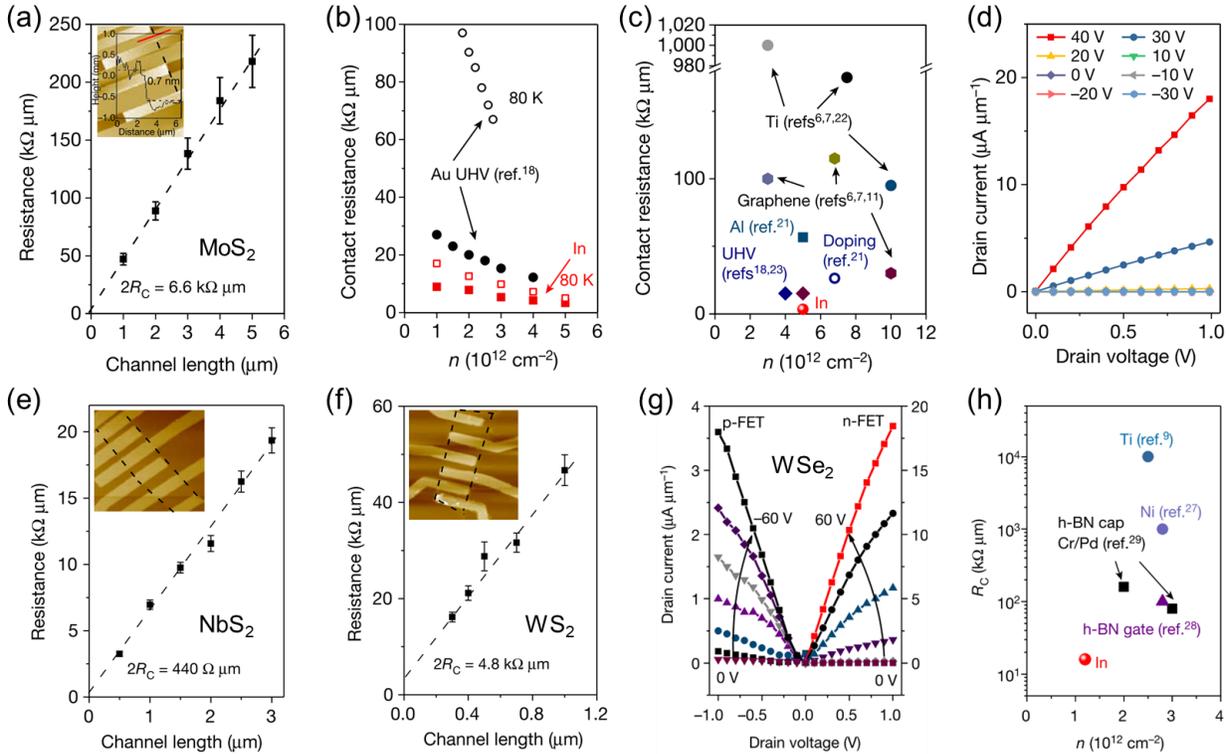

Figure 7. **Contact resistances and device properties of ultraclean In/Au electrodes on various 2D-TMDs**. (a) Contact resistance of the $MoS_2$-In/Au. (b) Contact resistance versus carrier concentration, $n$, for In/Au electrodes on $MoS_2$ at room temperature (filled points) and at 80K (open points). Comparisons are made with those of Au electrodes. (c) Comparing contact resistances with different types of electrode materials. (d) Linear output characteristics of $MoS_2$-In/Au contact indicates the absence of a contact barrier. (e) Contact resistance of In/Au electrodes on monolayer-$NbS_2$, and (f) monolayer-$WS_2$. (g) Linear output characteristics of ambipolar $WSe_2$-In/Au contact, and (h) comparing contact resistances with those of other electrode materials. Panels (a-h) reproduced from ref. [10] with permission from Springer Nature Limited, copyright 2019.

Meanwhile, it is important to note that device performances are often limited by the presence of interfacial residues when transferring TMD materials to a desired gate-dielectric substrate using conventional supporting holder polymethyl methacrylate (PMMA). To alleviate this problem, Mondal *et al*. proposed an approach to conduct residue-free wet-transfer via the use of polypropylene carbonate (PPC) as a supporting holder [87]. The fewer C=O bonds present in PPC compared to conventional PMMA result in weaker adsorption energy on the TMD surface. Consequently, PPC-transferred samples achieve ultra-clean, residue-free surfaces with a uniform distribution of TMD at the interface, leading to an ultralow ohmic contact resistance of approximately 78 Ω·μm and a high on/off ratio of ~$10^{11}$ at 15 K in FET devices. Furthermore, Kong *et al.* have utilized a thermally decomposable PPC as the buffer layer to realize an atomically clean and sharp vdW contact between metals and 2D- TMD, as indicated by various microscopy characterizations [72].

To fabricate a robust electrical contact and at the same time minimizing the effects of sample degradation and disorder, the problem of atmospheric oxidation arises as a challenge especially for air-sensitive 2D-TMDs [88]. Strategies to address this challenge have been proposed. For instance, Telford *et al.* fabricated a hole in hexagonal boron nitride using electron beam lithography and reactive ion etching, subsequently patterning a metallic contact that filled the cavity [89]. This marks the pioneering achievement of embedding metal contacts on insulating hexagonal boron nitride flakes, which are recognized for their exceptional encapsulating properties [89]. The hBN contraption can next be laminated onto 2D sheets such as $NbSe_2$, $MoTe_2$ and graphene as displayed in Figures 8a-c. Without the lithographic patterning of the metal contacts, this technique can minimize the problem of residual and structural damage at the metal-TMD contacts (Figures 8d-e). In addition, results from this study have demonstrated a low contact resistance in the order of ~$10^2$ Ω·μm for layered samples. Moreover, the air-sensitive monolayer-$NbSe_2$ maintains a low normal-state resistance with a superconducting transition temperature of ~3.7 K, an indication of its clean vdW contact and high device quality, that remains consistent over a prolonged period [90,91].

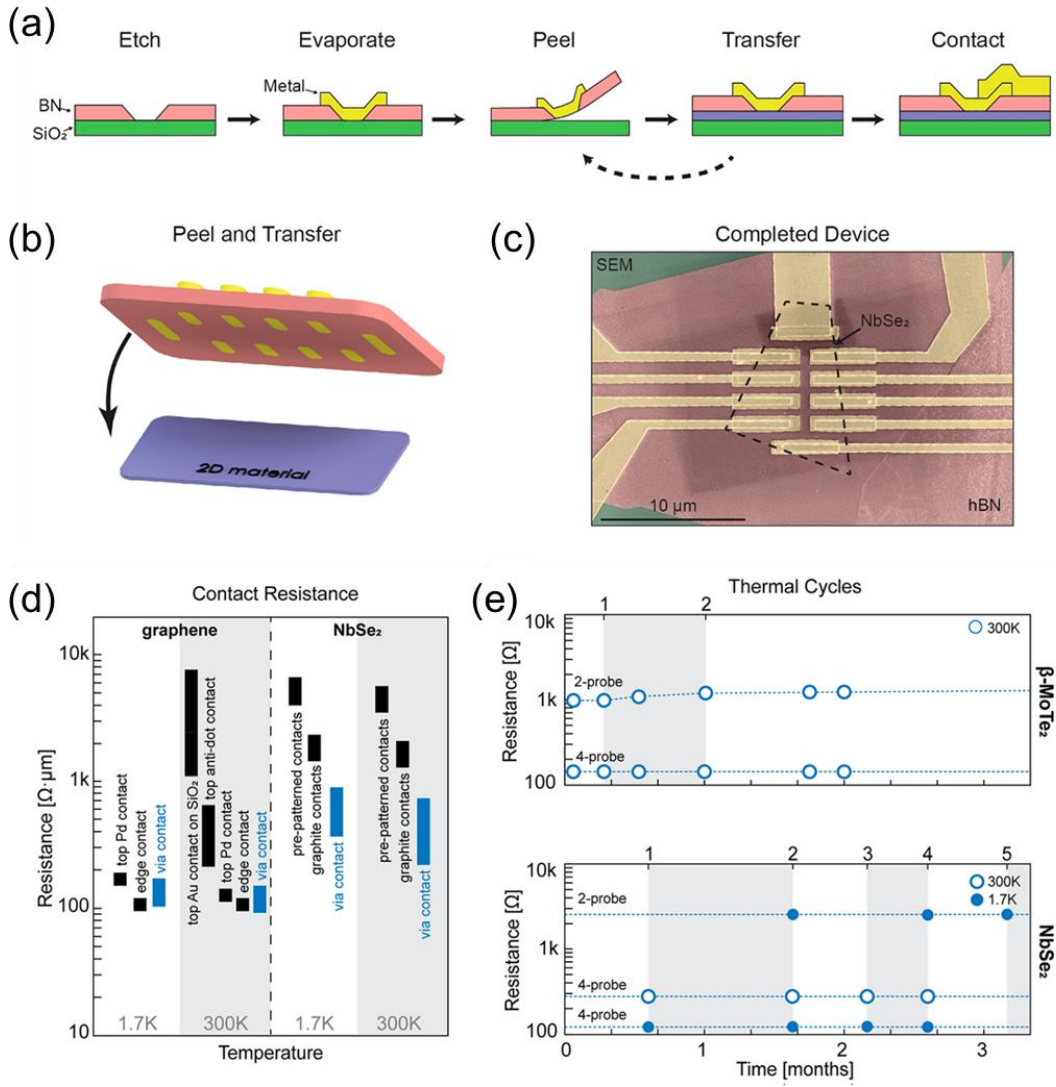

Figure 8. (a) Contact fabrication process where metal contacts are embedded into insulating hexagonal boron nitride flakes and thereafter laminated onto 2D-TMD sheets. (b) Positioning contact on 2D-TMD sheet, with (c) SEM imaging of the NbSe$_2$ contact system as an example. (d) Comparing contact resistances of the NbSe$_2$ system with other reported techniques. (e) 2- and 4-probe normal-state resistances of contacted NbSe$_2$ (top) and MoTe$_2$ (bottom) over time and thermocycles. Multiple configurations have shown stable resistances over time. Panels (a)-(e) reproduced from ref. [92] with permission from American Chemical Society, copyright 2018.

### 3.3 Prefabricated Metal Transfer

The conventional technique of depositing metal contacts during lithographic process may inevitably introduce contaminants to the metal/2D-TMDs interface and this in turn will lead to physical damage and degradation to the TMDs surface. Such effects ultimately resulting in poor contacts and degraded device performance. Liu *et al*. demonstrated the transfer of 3D metal thin-films laminated onto few-layered MoS$_2$ flakes (Figure 9a) without any interfacial chemical bonding [61]. This creates a vdW interface that is free from chemical disorder and Fermi-level pinning illustrated by TEM imaging of the interfacial system (Figures 9b-e). With the absence of these external experimental complications, the Schottky barrier of this contact can effectively be predicted by the Schottky-Mott rule that is free from any pinning effect [93,94]. However, the scalability of this technique has been restricted by the need for final alignment lithography to expose the contact under the thermoplastic polymer. To circumvent this issue, Went *et al*. developed the technique of metal contact transfer where the lithographic patterning procedures are conducted on a separate donor substrate instead of on the device itself (Figures 10a-b) [95]. Based on this technique, vertical Schottky-junction WS$_2$ solar cells fabricated with Ag and Au metals as asymmetric work function contacts yield a rectifying behavior with an open-circuit voltage above 500mV (Figure 10c). This is in contrast with the resistive behavior in devices fabricated via evaporated contacts.

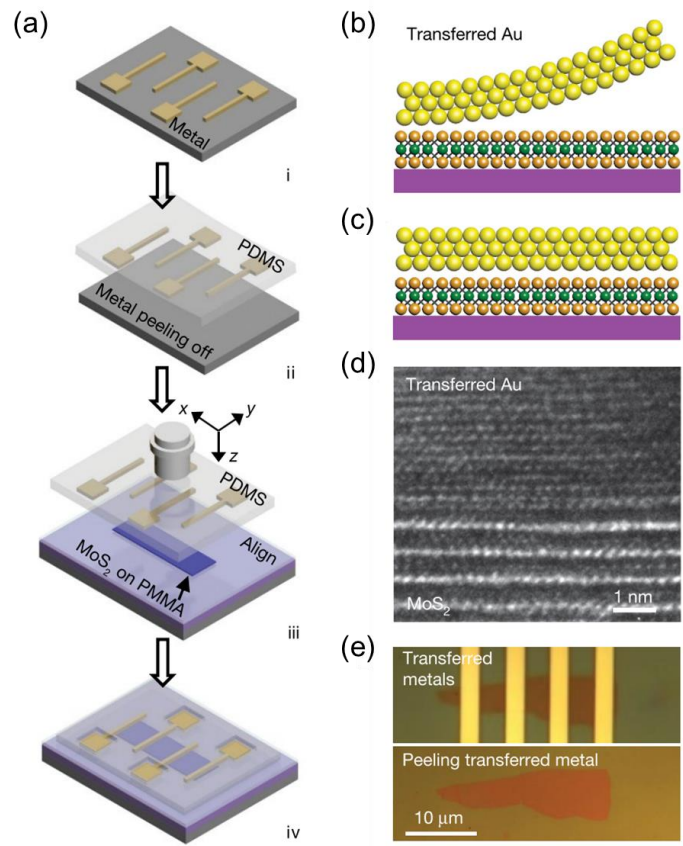

Figure 9. Illustration and structural characterizations of vdW metal–MoS$_2$ junctions. (a) Schematic depicting the transfer of 3D metal thin-films laminated onto few-layered MoS$_2$ flakes. (b-c) Cross-section schematics of transferred Au electrode above the MoS$_2$ flake, with (d) TEM imaging indicating an atomically sharp and clean metal-semiconductor interfaces. (e) Optical images of the MoS$_2$-Au device with, above: the transferred electrodes and, below: with the transferred electrodes mechanically released. Panels (a)-(e) reproduced from ref. [61] with permission from Springer Nature Limited, copyright 2018.

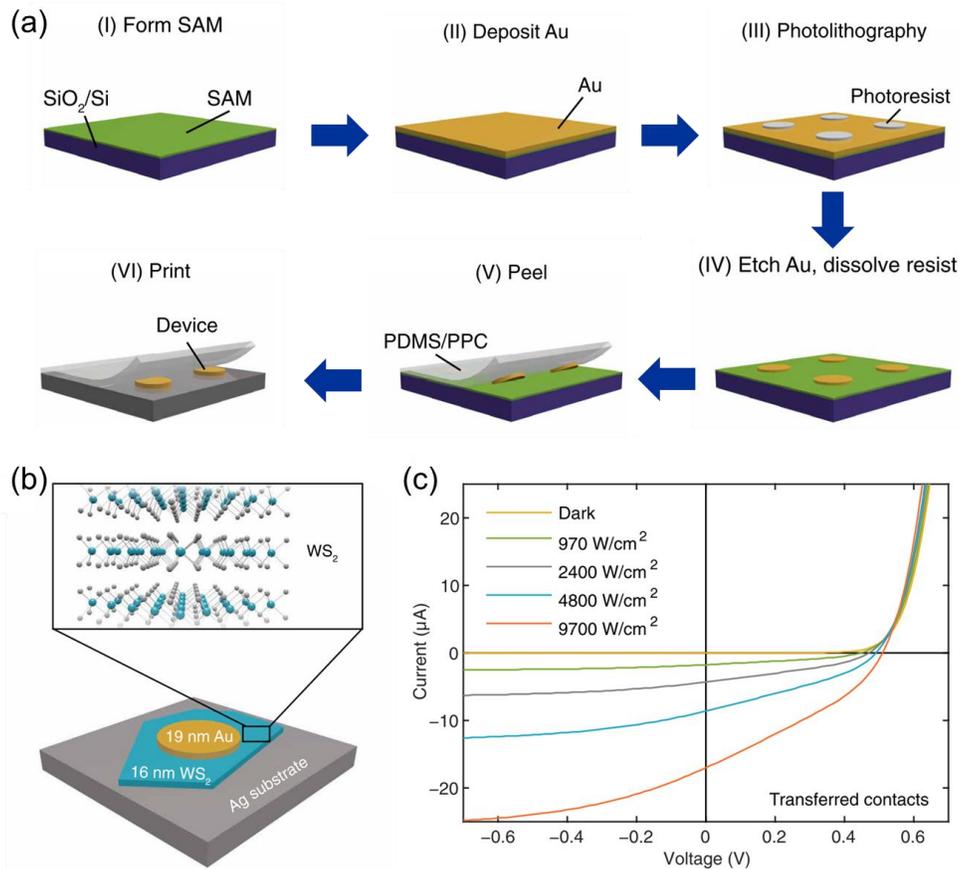

Figure 10. Schematics of the metal transfer process of the Schottky-junction multilayered WS$_2$ solar cells with transferred contacts. (a) Procedure of the metal transfer process. (b) schematic of the device structure with a 3D-representation of the atomic layers. (c) Power-dependent I-V characteristics of the device that display a rectifying behavior with an open-circuit voltage above 500 mV. Panels (a)-(c) reproduced from ref. [95] with permission from American Association for the Advancement of Science, copyright 2019.

Kong *et al.* utilized a prefabricated Au electrode pair, physically laminated it onto the WSe$_2$ flake [23]. Compared to conventional methods such as electron beam lithography and high vacuum thermal deposition, alternative metal transfer techniques offer cleaner Au/WSe$_2$ interfaces and enable control over Fermi level pinning or de-pinning effects. In a study by Liu *et al.*, prefabricated Ag/Au metal electrodes were mechanically laminated and transferred onto MoS$_2$/graphene vertical heterostructures [96]. This approach resulted in vdW metal contacts and short-channel vertical field-effect transistors with high on/off ratios. Microstructural analysis and transport studies demonstrated reduced direct tunneling current and Fermi-level pinning effects in this system. In addition, many efforts have been put into universal electrode transfer technologies. Liu *et al.* introduced a metal transfer printing process assisted by graphene, enabling the establishment of vdW contacts between 2D materials and 3D metal electrodes [97]. This graphene-assisted metal transfer printing approach facilitates the fabrication of MoS$_2$-based FETs with various printed metal electrodes, allowing for the tuning of Schottky barrier height and the formation of both ohmic and Schottky contacts.

Kong *et al.* demonstrated a wafer-scale universal electrode transfer technique for contacting semiconductors by using polymer buffer layers. These thermally decomposable polymers protect atomically thin semiconductors during metal vapor deposition and can be dry-removed through thermal annealing, resulting in vdW contacted 2D semiconductor devices. However, this method is not compatible with standard photolithography processes due to the polymer's solubility in the developer [72]. Recently, Xing et al. introduced a pick-and-place transfer technique for pre-fabricated electrodes from reusable polished hydrogenated diamond substrates without the need for surface treatments or sacrificial layers. This technique enables the transfer of large-scale, arbitrary metal electrodes with work functions ranging from 4.2 to 5.6 eV, allowing for the formation of low Schottky barriers by using suitable metals for monolayer transition metal dichalcogenides (TMDs). This technology shows significant promise for wafer-scale fabrication, leveraging advancements in large-area diamond disks [98].

### 3.4 Charge-transfer doping

Generally, 2D-TMDs contact with metals to fabricate FETs with n-type or p-type doping. High work function TMDs (such as $MoS_2$) favor n-type doping (Figures 11a-b), while low work function TMDs (such as $WSe_2$) prefer p-type doping to allow the ease of electron-transfer to flow the metal [39]. FLP effect at metal/2D-TMD interfaces tend to pin the fermi level of metal, approaching the valence band of TMDs [12]. This results in n-type devices being most commonly fabricated. N-type electrical contacts have a number of ways to overcome contact resistance, as mentioned earlier. Whereas in the cases of p-type systems including certain classes of transistors and light-emitting diodes where electron conduction is less desirable or where hole conduction is advantageous, the challenge remains to obtain low contact-resistance p-type electrical contacts. An alternative strategy to counter this problem is by doping the semiconductor-metal contacts to reduce the SBH to better facilitate the efficient quantum mechanical tunneling of carriers through the interface. While this approach has proven effective for establishing ohmic contacts in conventional Si-based MOSFETs, it poses challenges in precisely controlling and delineating doping processes within 2D semiconducting systems [99,100].

Surface charge transfer doping method has emerged as one of the most effective solutions to bypass the aforementioned limitations to create p-type ohmic contacts of 2D-TMD based systems even at cryogenic regime [101]. This technique relies on charge transfer between the 2D film and its surroundings through media such as surface adsorbates, adlayers/dielectric overlayers, substrates [102,103] where the exposed basal plane of 2D materials provides a favorable environment for dopants to adhere to the surface via chemical bonding or physisorption, thereby facilitating the doping process. By facilitating interfacial charge transfer while minimizing the introduction of substantial defects into the lattice structure of the as-doped materials, thereby largely preserving their fundamental transport characteristics [104]. Unlike conventional ion-implantation process which inflicts defects to the 2D crystal lattice, surface charge transfer doping with its selective area doping and mild treatment does not create any disorder in the film while the difference in the electron affinity between the dopant and the semiconducting 2D-TMD layer results in an injection or withdrawal of charge carriers to control the system's electrical conductivity (Figure 11c).

Based on this premise, Wang *et al* reported evaporation of high work function metal (Pt and Pd)

onto 2D TMDs to obtain clean vdW contacts, which effectively relieves the FLP effect at the metal—2D-TMDs interface and achieved purely p-type FETs [105]. This kind of p-type FET possess excellent output characteristics with saturation hole current of approximately 4 µA µm$^{-1}$ Figures 11d-e). In addition, a new approach, proposed by Chen *et al.*, employed high electron-affinity transition metal oxides, $MoO_3$, to degenerately dope the monolayer $WSe_2$ beneath electrodes, then form the Ohmic behavior monolayer $WSe_2$ FET down to 10 K [24]. Later, Xie *et al* [106], utilizes heavy hole-transfer doping and highly conductivity in the $WSe_2/\alpha$-$RuCl_3$ heterostructure to fabricate high performance p-type $WSe_2$ FETs. This is because $WSe_2$ has a significantly higher valence band maximum as compared to the conduction band minimum of $\alpha$-$RuCl_3$ (as inset of Figure 11f). This provides a favorable condition for achieving low-resistance p-type $WSe_2$ transistors. This approach reported a drain saturation current at on-state of up to 35 µA µm$^{-1}$, and an on-off ratio of ~$10^9$ at room temperature (Figures 11f-g). Another report by Das *et al.* [107] involves the substitutional doping of transition metal with V, Nb, and Ta onto thick $MoSe_2$ and $WSe_2$ to achieve p-type 2D FETs. They ingeniously designed a novel p-type 2D FETs that consist of monolayer 2D-TMD channels and doped multilayer contact regions. This strategy displayed high on/off current ratio above ~$10^4$ and a low contact resistance of ~2-5 kΩ µm. In addition, metal Pd with high electronegativity can act as an electron acceptor, thereby resulting in strong hybridization at the metal-TMD interface.

Interlayer coupling between bilayer $WSe_2$ and the metal layer can also modify the interfacial electronic band structure while spin–orbit coupling can alter the polarity at the $WSe_2$–Au contact [71,108]. All of them enable the metal-TMD system to be p-type.

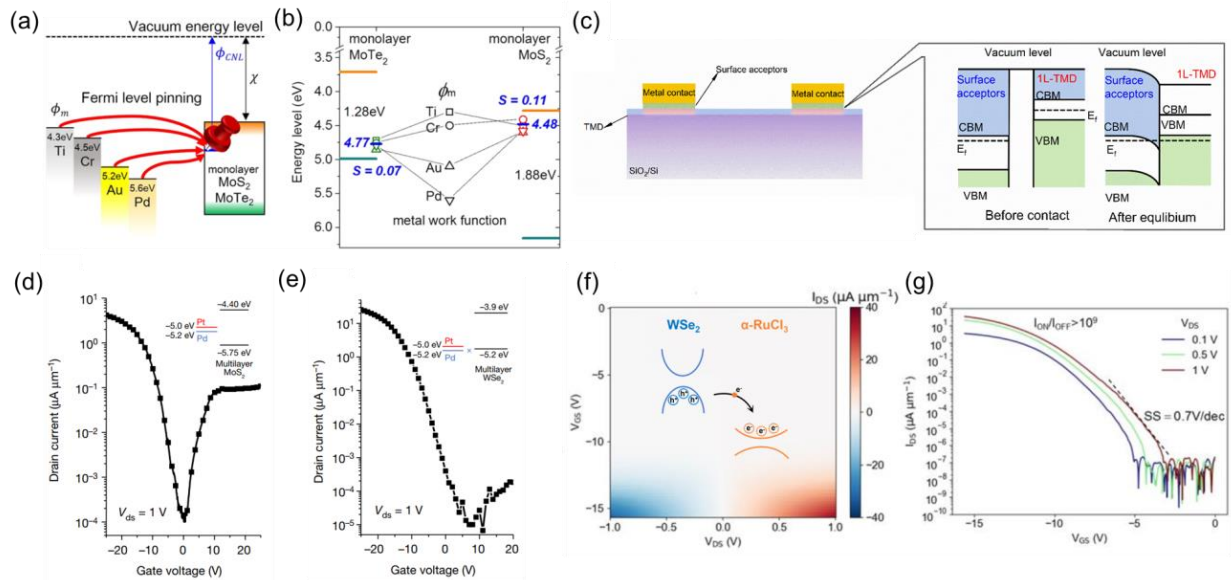

Figure 11. (a) Energy band structure of various metals and monolayer TMDs. (b) Diagram of FLP effect at metal–MoS$_2$/TMDs interfaces. (c) Schematic diagram of charge transfer process between high electron affinity surface acceptors and 2D materials. Drain-source current as a function of Gate-source voltage about (d) MoS$_2$- Pd and (e) WSe$_2$-Pt contacts presenting more excellent p-type characteristics compared to n-type. (f) Output mapping of monolayer WSe$_2$ FET show the maximum I$_{DS}$ (drain-source current) is 35 µA µm$^{-1}$. The inset is band alignment of WSe$_2$ and α-RuCl$_3$. The arrow points in the direction of the electron flow. (g) the ratio of on/off drain-source current at different drain-source voltage. Panels (a)-(b) reproduced from ref. [39] with permission from American Chemical Society, copyright 2017. Panels (d)-(e) reproduced from ref. [105] with permission from Springer Nature Limited, copyright 2022. Panels (f)-(g) reproduced from ref. [106] with permission from American Chemical Society, copyright 2024.

## 4. Applications
### 4.1 Electronic devices

2D-TMDs materials have led to extensive applications in electronic devices due to their unique and excellent electronic properties. Examples of electronic devices are diverse and they include diodes, field-effect transistors, integrated circuits, nonvolatile memories [105,109,110], etc. To improve the device performance and optimize their efficiency, it is essential to minimize contact resistance so as to meet the requirements of new-generation electronics. A novel method has been proposed by Sheng *et al.* in the fabrication of 2D-TMD-based transistors that uses soft plasma treatment to remove sulfur end atoms and to directly link the middle Mo atoms and metal electrode [109]. Removing the sulfur atoms by soft plasma treatment changes the band structure of TMDs and transforms the original metal-semiconductor contact into a metal-metal contact, resulting in the improved performance of the fabricated FETs. Kim *et al.* have also demonstrated that the introduction of a ZnO nanowire gate FET displays a good ON/OFF current ratio [110]. Specifically, long 1D-ZnO nanowires have been utilized as gate electrodes to fabricate 2D-TMD channel field-effect transistors and 2D-TMD complementary nonvolatile memories. Wang *et al.* reported high-performance field-effect transistors with diode properties based on electron beam evaporation technology of high-work-function metals [105]. This electron beam evaporation technology creates non-interfacial-defect vdW contacts between high-work-function metals and 2D-TMDs. The essential properties including the power output, power conversion efficiency, responsivity and external quantum efficiencies for $MoS_2$ and $WSe_2$ devices are amongst the highest reported [105].

**4.2 Spintronic devices**

There is an imperative for the advancement of novel engineering methodologies pertinent to devices incorporating two-dimensional transition metal dichalcogenides (2D-TMDs). Notably, within the realm of spintronic applications, the integration of a graphene-TMD bilayer structure emerges as a topic of considerable interest. The strategic assembly of such a configuration is motivated by the advantageous positioning of the graphene Dirac cone within the semiconducting bandgap of the TMD material. This spatial arrangement ensures the persistence of the $\pi$-band, concurrently facilitating the robust incorporation of spin-orbit coupling effects emanating from the TMD layer [111-113]. Dankert and Dash demonstrated the electrical gate tunability of the spin current and lifetime of a graphene-$MoS_2$ vdW heterostructure at room temperature [114]. It was further shown that the spin degree-of-freedom in the graphene layer can interact and de-phase efficiently in the $MoS_2$ layer via the gate-voltage tuning of the Schottky barrier at the graphene–$MoS_2$ interface. Benitez *et al.* then showed that anisotropic spin dynamics can take place with $WS_2$ or $MoS_2$ in the graphene-TMD heterostructure [115,116]. There is a change in spin lifetimes by over an order of magnitude between the in- and out-of-plane spins – out-of-plane spins being the larger. This observation indicates the presence of strong spin-valley coupling of the TMD layer is imprinted within the heterostructure which in turn influences the propagating spins [117]. Similar observations have also been made in a separate study by Ghiasi *et al.* in the $MoSe_2$-graphene heterostructure showing 1 order of magnitude longer lifetime for the out-of-plane spins ($\tau_\perp \approx 40$ ps) as compared to its in-plane spins ($\tau_\parallel \approx 3.5$ ps) [118]. Therefore, spintronics as well as spintronic devices would become increasingly important for TMDs materials.

### 4.3 Optoelectronic devices

Beyond their remarkable electronic and spin properties, the optical characteristics of 2D-TMDs, particularly exemplified by their monolayer-level thickness, are of profound interest for optoelectronic applications, including photodetectors [119,120]. The allure of TMDs as optoelectronic materials stems from their inherent transparency, mechanical flexibility, wide bandgap choices, and facile material processing capabilities. Manipulating the number of TMD layers, can also modulate the bandgap, rendering them effective light detectors across diverse wavelengths [121,122]. Furthermore, metallic and quasi-metallic phase TMDs exhibit notably elevated photocurrents compared to their semiconducting counterparts, thereby finding wide-ranging utility in various optoelectronic device implementations [123,124]. The work by Yamaguchi *et al*. involves a series of experimental techniques including scanning photocurrent microscopy (SPCM) and photoluminescence imaging[125]. They show that the H-T phase transition of $MoS_2$ leads to a strong reduction in the Schottky barrier. The metallic phase $MoS_2$ device also registered an enhancement in photoresponsivity by more than an order of magnitude.

## 5. Summary and outlook

### 5.1 Summary

In this review, we have reviewed the different cases for metal contacts with conventional semiconductors and two-dimensional (2D) transition metal dichalcogenides (TMDs). Proper selection of a metal electrode with a matched work function is a simple and effective method that could lower the Schottky barrier in metal-conventional semiconductor contacts. However, the properties of contacts to 2D-TMD materials cannot be intuitively predicted by metal work function values. Previous research has indicated that FLP is the main bottleneck to overcoming contact resistance in metal/2D-TMDs interfaces due to interfacial defects, metal induced gap state (MIGS) and chemical distortion. Therefore, the key to overcoming contact resistance is to eliminate or weaken the Fermi level pinning in metal–2D-TMDs contacts. Numerous engineering strategies have been established to deal with the challenges of Fermi level pinning in metal–2D-TMDs contacts. A Fermi level pinning-free contact can be realized through interfacial hybridization engineering, vdW contact, metal mechanical transfer and charge-transfer doping. Large size of the atomic orbitals, small physical separations, chalcogen vacancies, chalcogen interstitial doping and edge contacts can enable strong orbital hybridization at metal/2D-TMDs interfaces, which narrows the Schottky barrier. Insertion of a buffer layer can eliminate MIGS and facilitate carrier injection in the moderate degree of interfacial hybridization. Prefabricated metal transfer as well as ultraclean vdW contact can also boost the device performance at the nanoscale. Based on the fragile nature of this set of materials, surface-transfer-doping method has been demonstrated as an effective way to form p-type ohmic contacts for TMD materials.

## 5.2 Outlook

Despite the promise and growing prospects of metal–2D-TMD contacts, numerous challenges persist in optimizing their properties for specific applications. Achieving low contact resistance and efficient charge carrier injection remains a formidable task, requiring intricate engineering solutions and precise control over interface characteristics. Nevertheless, considerable efforts have been dedicated to identifying suitable metal materials for forming ohmic contacts, employing both theoretical calculations and experimental findings. These numerous research efforts have the primary aim to minimizing and overcoming contact resistance in 2D-TMD-based devices which hinders advancement. Understanding the mechanisms of Fermi level pinning at metal-2D-TMD interfaces is essential, necessitating further scientific elucidation. Moreover, strategies aimed at mitigating contact resistance encounter many obstacles. Predictions from interfacial hybridization studies suggest that robust hybridization between metals and TMDs can metallize the underlying TMDs and abolish tunneling barriers, thereby reducing contact resistance. However, real-world conditions present challenges, as the atomic-scale thickness and pristine surfaces of TMDs render them susceptible to environmental surface adsorbates, thereby constraining hybridization [39]. While the insertion of a buffer layer can alter the interface between TMDs and metal, its effectiveness in reducing contact resistance is limited. Extending interfacial hybridization engineering to ultrathin 2D semiconductor devices remains an open question [126]. Enhancing the interfacial purity of metal–2D-TMD during sample preparation and contact assembly is imperative for future progress.

Further research is needed to overcome challenges such as strong Fermi level pinning and high contact resistance in interfacial hybridization engineering. Successfully reducing contact resistance in metal–2D-TMD contacts is crucial for achieving high device performance.

On a longer time scale, it is imperative to take into consideration the eventual goal of developing and integrating these fundamental techniques with advanced manufacturing processes and scaling them up for future advanced electronic, optoelectronic, energy, and emerging technologies [127]. While it appears to be a generalization of upscaling and refining contact resistances in the realm of 2D-TMD based systems, intricate details concerning the fundamental vdW mechanisms, device engineering, manufacturing technologies and methodologies ought to be taken into collective consideration in advancing this discipline and eventually actualizing it in practical device applications. Therefore, the design and fabrication of optimized metal–2D-TMD contacts tailored for specific applications necessitate the precise control of contact resistance and efficient charge carrier injection [128]. The proper management of defects from the molecular to the device scale and the importance of ensuring device and contact structural integrity are paramount to ensuring device reliability and durability.

Scalability poses another significant hurdle to be appropriately addressed. While successes have been achieved in minimizing contact resistance in laboratory settings, the primary challenge remains in the transition from laboratory-scale research to large-scale manufacturing [27]. Thus, this upscaling process demands innovative approaches and technological advances which may not be sufficiently mature or may not even exist the present. With the development of new technologies and engineering techniques for the upscaling of device fabrication processes, it is equally important to ensure the sustained reliability, durability and robustness of TMD-based devices and contact systems under real-world conditions. This demands comprehensive characterization and novel materials engineering strategies to ensure the feasibility of implementing TMD-based devices in our lives where technology plays a ubiquitous role. The basis of the aforementioned considerations lies with the standardization of protocols for TMD material synthesis, device fabrication, and performance assessment. These factors will ensure consistency and reliability across various research domains and industrial applications [19]. Technological challenges also necessitate interdisciplinary collaboration and innovative device design approaches.

In conclusion, amidst the concurrent advancement of materials science and engineering techniques, it is important to foster of interdisciplinary cooperation. This is because such a collective approach holds the key to unlocking the full potential of TMD-based devices and driving their widespread adoption across diverse applications. While the challenges prevail, the immense potential of TMD-based devices and systems vis-à-vis vdW contacts continues to fuel global research and development efforts. With these engineering challenges, by seizing upon new opportunities for innovation, this burgeoning field of 2D-TMDs and metal contacts is poised to make significant contributions to the advances in electronic, optoelectronic, energy, and emerging technologies in the years and decades ahead.

Last but not least, the exquisite electronic, spin and optical properties of 2D-TMDs materials as well as realistic devices based on 2D-TMDs materials have emerged. Although there are still some unsolved challenges in this field of contact resistance, we confidently anticipate that 2D-TMDs will become an important new material for advanced applications.

## Conflicts of interest

There are no conflicts to declare.

## Acknowledgement

This work was supported by National Natural Science Foundation of China (Grant No. 52172271, 12374378, 52307026), the National Key R&D Program of China (Grant No. 2022YFE03150200), Shanghai Science and Technology Innovation Program (Grant No. 22511100200, 23511101600). X. L acknowledges the support from the China Scholarship Council (CSC) (Grant No. 202306890069). C.S.T acknowledges the support from the NUS Emerging Scientist Fellowship. S.S. acknowledges support from Natural Science Foundation of China (Grant No. 12304199), Science and Technology Commission of Shanghai Municipality, the Shanghai Venus Sailing Program (Grant No. 23YF1412600). A. T. S. W acknowledges the funding support by the Ministry of Education, Singapore, under MOE Tier 2 grant MOE-T2EP50122-0020.